\documentclass[12pt]{article}
\usepackage{graphics}
\usepackage{amssymb}
\usepackage{psfrag}

\newcommand{\rf}[1]{(\ref{#1})}
\newcommand{\beq}{\begin{equation}}
\newcommand{\eeq}{\end{equation}}
\newcommand{\bea}{\begin{eqnarray}}
\newcommand{\eea}{\end{eqnarray}}

\newcommand{\lam}{\lambda}
\newcommand{\La}{\Lambda}

\renewcommand{\a}{\alpha}

\newcommand{\om}{\omega}
\newcommand{\del}{\delta}
\newcommand{\Del}{\Delta}

\newcommand{\oh}{\frac{1}{2}}

\newcommand{\ra}{\rangle}
\newcommand{\la}{\langle}

\newcommand{\equ}{\!=\!}

\newcommand{\cD}{{\cal D}}

\begin{document}

{\normalsize \hfill SPIN-2004/22}\\
\vspace{-1.5cm}
{\normalsize \hfill ITP-UU-04/40}\\
${}$\\ 

\begin{center}
\vspace{48pt}
{ \Large \bf  Semiclassical Universe from First Principles}

\vspace{40pt}

{\sl J. Ambj\o rn}$\,^{a,c}$,
{\sl J. Jurkiewicz}$\,^{b}$
and {\sl R. Loll}$\,^{c}$

\vspace{24pt}
{\footnotesize

$^a$~The Niels Bohr Institute, Copenhagen University\\
Blegdamsvej 17, DK-2100 Copenhagen \O , Denmark.\\
{ email: ambjorn@nbi.dk}\\

\vspace{10pt}

$^b$~Mark Kac Complex Systems Research Centre,\\
Marian Smoluchowski Institute of Physics, Jagellonian University,\\ 
Reymonta 4, PL 30-059 Krakow, Poland.\\
{email: jurkiewicz@th.if.uj.edu.pl}\\

\vspace{10pt}

$^c$~Institute for Theoretical Physics, Utrecht University, \\
Leuvenlaan 4, NL-3584 CE Utrecht, The Netherlands.\\ 
{email:  j.ambjorn@phys.uu.nl, r.loll@phys.uu.nl}\\

\vspace{10pt}

}
\vspace{48pt}

\end{center}


\begin{center}
{\bf Abstract}
\end{center}

Causal Dynamical Triangulations in four dimensions provide
a background-independent definition of the sum over space-time
geometries in nonperturbative quantum gravity. We show
that the macroscopic four-dimensional world which emerges 
in the Euclidean sector of this theory is a bounce which satisfies  
a semiclassical equation. After integrating out all degrees of freedom 
except for a global scale factor, we obtain the ground state wave 
function of the universe as a function of this scale factor.

\vspace{12pt}
\noindent


\newpage

\subsection*{1. Introduction}\label{intro}

One important application of any theory of quantum gravity is a
description of the quantum evolution of the very early universe. 
This is also the realm of quantum cosmology, which
tries to capture the essence of the gravitational dynamics
by quantizing only a finite number of degrees of
freedom characterizing the universe as a whole. The path integral
formulation of quantum cosmology came to 
prominence with the work of the Cambridge group and others on
Euclidean quantum gravity \cite{eqg}, and in particular that of 
Hartle and Hawking \cite{hh}.    
Central in this and related approaches is the construction of a
``wave function of the universe", either as a solution of the
Wheeler-DeWitt equation or a propagator for the theory (see, for example,
\cite{vilenkin,linde,rubakov,hl,haha}). In
attempting to do this, a variety of technical and conceptual 
issues has to be addressed, including the choice
of boundary conditions for the wave function, the unboundedness
of the gravitational action and ensuing divergence of the
Euclidean cosmological path integral, the appropriateness of
the minisuperspace and/or semiclassical approximations, and the physical 
interpretation of the construction (see \cite{vilrev} for a recent 
concise review). 

One could hope that a nonperturbative path
integral formulation which does not impose any a priori symmetry
restrictions on the geometry of the universe
would help resolve some of these issues. ``Causal 
Dynamical Triangulations'' provide 
exactly such a background-independent, nonperturbative 
definition of quantum gravity, in which the 
sum over all space-time geometries is constructively defined
and the causal (Lorentzian) structure of space-time plays a
crucial role \cite{al,ajl3d,ajl-prd,ajl4d,ajl-prl}. 
It can be viewed as a realization of an idea of Teitelboim's, who
argued that in a (continuum) proper-time formulation of the Lorentzian
gravitational path integral one should integrate over positive 
lapse functions only, thereby building a notion of causality
into the quantum dynamics \cite{teitelboim}. 

In the context of quantum cosmology it has been argued 
\cite{hl,vilcos} that a tunneling wave function \`a la Vilenkin, 
where a universe tunnels from a vanishing 
to an extended three-geometry, is a special case of Teitelboim's 
causal propagator between two three-geometries. In the present work, 
where the wave function of the universe will be constructed from first 
principles, we will indeed observe a similar phenomenon,
although our interpretation in the end will be 
somewhat different.

In \cite{ajl-prl} we reported that the approach of Causal 
Dynamical Triangulations, 
despite its background independence, generates a four-dimensional 
universe
around which (small) quantum fluctuations take place.
The purpose of this letter is to identify the effective 
action which determines the shape of this macroscopic 4d world.
Rather surprisingly we find that the effective action
which describes the infrared, long-distance part of the 
universe is closely related to a simple minisuperspace
action frequently considered in quantum cosmology. 
The only differences in our full quantum treatment
are that (i) the unboundedness problem of the 
conformal mode in the Euclidean sector is cured, and (ii) 
the ultraviolet, short-distance part of the effective action is 
such that the solution to the Euclidean action describes
a bounce from a universe of no spatial extension to one 
of finite spatial size. While resembling Vilenkin's picture of a
``universe from nothing" \cite{vilenkin}, the interpretation in the 
present context is rather in terms of the ground state wave 
function of the universe with everything but the scale factor 
integrated out. We describe how one can determine this 
wave function from first principles.

It should be emphasized that unlike in standard minisuperspace
models for cosmology, we do not assume homogeneity or
isotropy, nor do we impose any other a priori symmetry conditions
on the gravitational
degrees of freedom. We perform the full path integral and determine
the effective Lagrangian which describes the dynamics of the
global scale factor, as well as the ground state wave function 
of the universe as a function of this scale parameter.

The rest of this article is organized as follows: in the next 
section we recall some salient features of the Causal Dynamical 
Triangulations approach, including the set-up 
of the numerical simulations and recent numerical results in
four dimensions.
We then demonstrate in Sec.\ 3 that the numerical data are perfectly
described by a simple minisuperspace action. Sec.\ 4 
outlines how this result relates to the ground state wave function 
of the universe, and the final Sec.\ 5 is devoted to a discussion.

\subsection*{2. Observing the bounce}
 
The idea to construct a quantum theory of gravity by using Causal
Dynamical Triangulations was motivated by the desire to 
formulate a quantum gravity theory with the correct Lorentzian 
signature and causal properties \cite{teitelboim}, and to
have a path integral formulation which may be closely related 
to attempts to quantize the theory canonically. 
For the purposes of this letter, we will only summarize the main
properties of this approach; more details on the rationale and techniques
can be found elsewhere \cite{al,ajl3d,ajl-prd,ajl4d}.

We insist that only causally well-behaved geometries appear in
the path integral, which is regularized by summing over a particular
class of triangulated, piecewise flat (i.e.\ piecewise Minkowskian)
geometries. All causal simplicial space-times contributing to the path integral
are foliated by a version of ``proper time" $t$, and each geometry
can be obtained by gluing together four-simplices 
in a way that respects this foliation. Each four-simplex has time-like
links of negative length-squared $-a_t^2$ and space-like links
of positive length-squared $a_s^2$, with all of the latter located in spatial
slices of constant (integer in lattice units) proper time $t$.
These slices consist of purely
space-like tetrahedra, forming a three-dimensional piecewise flat
manifold, whose topology we choose for simplicity 
to be that of a three-sphere
$S^3$. A necessary condition for obtaining a well-defined 
continuum limit from this regularized setting is that
the lattice spacing $a\propto a_t\propto a_s$ goes to zero while the 
number $N_4$ of four-simplices goes to infinity in such a way that the
continuum four-volume $V_4:= a^4N_4$ stays fixed. 
Let us emphasize that the parameter $a$ therefore does not play
the role of a fundamental discrete length.
A further property of our explicit construction is that each
configuration can be rotated to Euclidean signature, a
necessary prerequisite for discussing the convergence properties
of the sum over geometries, as well as for using Monte Carlo
techniques.

The partition function for quantum gravity is
\beq\label{2.1}
Z(\La,G) = \int \cD [g] \; e^{iS[g]},~~~
S[g]= \frac{1}{G}\int d^4x \sqrt{|\det g|}\,(R-2\La),
\eeq
where $S[g]$ is the Einstein-Hilbert action including a 
cosmological-constant term $\La$, and $G$ the gravitational 
constant\footnote{We ignore a numerical constant
multiplying $G$ in \rf{2.1}.}.
Using our simplicial regularization this becomes
\beq\label{2.2}
Z(\La,G)_{CDT} = \sum_T \frac{1}{C_T} \, e^{iS[T]},
\eeq
where the integration over Lorentzian geometries is replaced by
a sum over causal triangulations,
and $C_T$ is a symmetry factor of the triangulation T, the
order of its automorphism group.
The action $S[T]$ is  Regge's version of the
Einstein-Hilbert action, appropriate for piecewise linear geometries
(see \cite{ajl4d} for details).
In the remainder of this article we will for simplicity use a continuum 
notation, but it should be understood that whenever computer
simulations are mentioned the implementation of the path integral is in
terms of piecewise flat geometries.

The substitution $- a_t^2 \to a_t^2$ turns all time-like into space-like
edges and rotates all configurations $g$ to Euclidean space-times
$g_E$, replacing at the same time $i S[g] \to -S_E[g_E]$, 
where $S_E$ denotes the
Euclidean Einstein-Hilbert action. In computer simulations it is often 
convenient to work with universes of constant four-volume $V_4$. 
The Euclidean counterpart of the partition function \rf{2.1} can be
decomposed as
\beq\label{2.1a}
Z_E(\La,G) = \int_0^\infty dV_4\; e^{-\frac{1}{G} \La V_4} 
\tilde Z_E(V_4,G),
\eeq
where the partition function $\tilde Z_E(V_4,G)$ for fixed four-volume is defined as
\beq\label{2.1b}
\tilde Z_E(V_4,G) =\! \int\!\cD [g] \, e^{-\tilde S_E[g]}\, \del(\! \int\! d^4x \sqrt{\det g}-V_4),~~~
\tilde S_E[g]= -\frac{1}{G}\!\int\! d^4 x \sqrt{\det g}\,R.
\eeq
Whenever $V_4$ is kept fixed
we will use $\tilde Z_E(V_4,G)$ as our partition function. It is related to
$Z_E(\La,G)$ by the Laplace transformation \rf{2.1a}. 

The specific (Euclidean) partition function we will consider is the
so-called quantum-gravitational proper-time propagator defined by
\beq\label{2.3}
G_{\La,G}^{E}( g_{3}(0),g_{3}(t)) = \int \cD [g_E] \; e^{-S_E[g_E]}.
\eeq
where the integration is over all four-dimensional (Euclidean)
geometries $g_E$ of topology $S^3\times [0,1]$, each with proper time
running from 0 to $t$, and with spatial
boundary geometries $g_{3}(0)$ and $g_{3}(t)$ at proper times $0$ and $t$.

\begin{figure}[ht]
\centerline{\scalebox{1.1}{\rotatebox{0}{\includegraphics{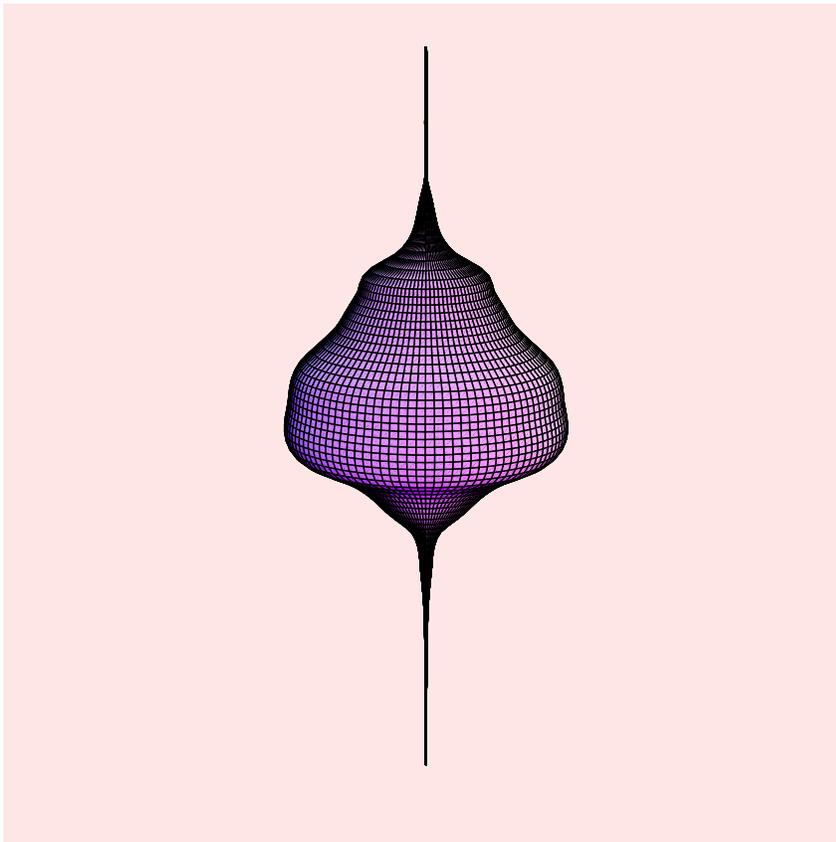}}}}
\caption[phased]{{\small
Monte Carlo snapshot of a ``typical universe'' of discrete volume 91.100 
four-simplices 
and total time extent (vertical direction) $t=40$. The circumference at 
integer proper time $s$ is proportional to the spatial three-volume 
$V_3(s)$. The surface represents an interpolation between
adjacent spatial volumes, without capturing the
actual 4d connectivity between neighbouring spatial slices.}}
\label{fig1}
\end{figure}
While it may be difficult to find an explicit analytic
expression for the full propagator \rf{2.3} of the four-dimensional theory,
Monte Carlo simulations are readily available, using standard
techniques from Euclidean dynamically triangulated quantum gravity \cite{MC}.
For convenience of the computer simulations, we keep the total
four-volume $V_4$ of space-time fixed and also often use
periodic rather than fixed boundary conditions, i.e. sum over 
space-times with topology $S^3\times S^1$ rather than
$S^3\times [0,1]$. This periodicity does not affect the results reported
below, as is illustrated by Fig.\ref{fig1}. This shows the typical ``shape"
(spatial three-volume $V_3(s)$ as a function of proper time $s$) of a 
space-time
configuration generated by the computer.\footnote{``Typical" in this context
is used to characterize the geometric characteristics that will be shared
with probability 1 by a randomly chosen member of the ensemble of
four-geometries.} For given space-time volume $V_4$, as long as $t$ 
is chosen sufficiently large, the configuration will develop a thin stalk
like the one shown in Fig.1. It will then not matter for the analysis of the
large-scale geometry whether or not time is periodically identified.

\begin{figure}[ht]
\vspace{-3cm}
\psfrag{d}{\bf{\Large $x$}}
\psfrag{VV}{\Large\bf $C_{V_4}(x)$}
\centerline{\scalebox{0.6}{\rotatebox{0}{\includegraphics{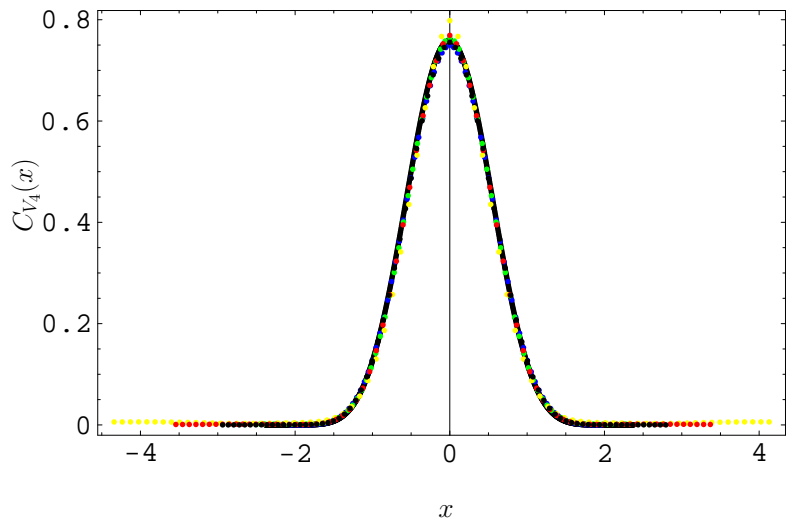}}}}
\vspace{-4.5cm}
\caption[phased]{{\small Measurement of spatial volume-volume correlator 
for space-times with 22.250, 45.500, 91.000, 181.000 and 
362.000 four-simplices, plotted as function of the scaled variable 
$x=\Delta/V_4^{1/4}$.}}
\label{fig2}
\end{figure}

A convenient ``observable'' is the spatial volume-volume correlator defined by
\beq\label{v-v}
C_{V_4}(\Delta)\equiv \la V_3(0)V_3(\Delta)\ra_{V_4} = 
\frac{1}{t^2} \int_0^t ds \;\la V_3(s)V_3(s+\Del)\ra_{V_4},
\eeq
where we have identified the spatial boundary geometries
at times 0 and $t$.
We have measured this correlator for various four-volumes $V_4$ and 
plotted the result as a function of the scaled variable 
$x=\Del/V_4^{1/d}$, after normalizing the correlator for a given 
$V_4$ to have an integral equal to 1 (in lattice units).
The variable $d$ is then determined from the condition that 
the overlap between the correlators of different $V_4$
be maximal.\footnote{We are 
employing standard finite size scaling methods from the 
theory of critical phenomena in statistical mechanics, see, for example \cite{bark}.}
Fig.\ref{fig2} illustrates the almost perfect overlap obtained for $d=4$.
We take this as strong evidence that our ``macroscopic'' space-times
(we are using up to 360.000 four-simplices) are genuinely 
four-dimensional.\footnote{Similar data, but with a smaller maximal
four-volume, were already published in \cite{ajl-prl}.}
As emphasized in \cite{ajl-prl}, this is a highly non-trivial result and
does {\it not} follow from the fact that the individual building blocks at 
the cut-off scale $a$ are four-dimensional.
Additional evidence for a macroscopically four-dimensional universe 
was presented in \cite{ajl-prl}, and an extended analysis will appear 
in due course \cite{more}. 

\begin{figure}[ht]
\vspace{-3cm}
\psfrag{x}{\bf{\Large $z$}}
\psfrag{T}{\Large\bf $P_{V_3}(z)$}
\centerline{\scalebox{0.6}{\rotatebox{0}{\includegraphics{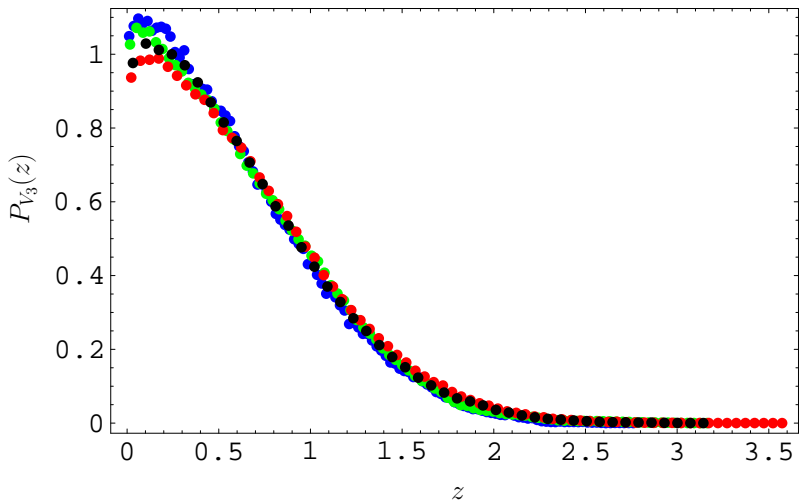}}}}
\vspace{-4.5cm}
\caption[phased]{{\small Distribution $P_{V_3}(z)$ of volume differences
of adjacent spatial slices, for three-volumes $V_3=$ 10.000, 20.000, 
40.000 and 80.000 tetrahedra.}}
\label{fig3}
\end{figure}

{\it Our goal here will be to 
understand the precise analytical form of the volume-volume correlator
$C_{V_4}(\Delta)$.} To this end, let us consider the 
distribution of differences in the spatial volumes $V_3$ of successive spatial slices 
at proper times $s$ and $s+\del$, 
where $\del$ is infinitesimal, i.e.\ $\del=1$ in lattice 
proper time units. We have measured the probability 
distribution $P_{V_3}(z)$ of the variable
\beq\label{z}
z= \frac{V_3(s+\del)-V_3(s)}{V_3^{1/2}},~~~~V_3=V_3(s)+V_3(s+\del).
\eeq
for different values of $V_3$. As shown in 
Fig.\ \ref{fig3} they fall on a common curve.\footnote{Again we have 
applied finite size scaling techniques, starting out
with an arbitrary power $V_3^{\a}$ in the denominator in \rf{z}, 
and have determined $\a =1/2$
from the principle of maximal overlap of the distributions for various 
$V_3$'s.} Furthermore, the distribution $P_{V_3}(z)$ is fitted very well by 
a Gaussian $e^{-c z^2}$, with a constant $c$ independent of $V_3$.   
From estimating the entropy of spatial geometries, that is, 
the number of such configurations,
one would expect corrections of the form  $V_3^\a$, with $0\leq \a < 1$, 
to the exponent $c \, z^2$ in the distribution $P_{V_3}(z)$. Unfortunately 
it is impossible to measure these corrections directly in a reliable way. 
We therefore make a general ansatz for the probability distribution 
for {\it large} $V_3(s)$ as
\beq\label{2.6}
\exp\left[- \frac{c_1}{V_3(s)} \left(\frac{dV_3(s)}{ds}\right)^2
-c_2 V_3^{\a}\right],
\eeq
where $0\leq \a <1$, and $c_1$ and $c_2$ are positive constants. 

We are thus by  ``observation'' led to  the following 
effective action for large three-volume $V_3(s)$:
\beq\label{2.7}
S^{(eff)}_{V_4} = \int_0^t ds \; 
\left(\frac{c_1}{V_3(s)} \left(\frac{dV_3(s)}{ds}\right)^2
+c_2 V_3^{\a}(s) - \lambda V_3(s)\right),   
\eeq
where $\lambda$ is a Lagrange multiplier to be determined 
such that
\beq\label{2.8}
\int_0^t ds \, V_3(s) = V_4.
\eeq
From general scaling of the above action it is clear that
the only chance to obtain the observed scaling law, expressed in terms of
the variable $t/V_4^{1/4}$, is by setting $\a= 1/3$. In addition, to reproduce
the observed stalk for large times $t$ the function $V_3^{1/3}$ has to
be replaced by a function of $V_3$ whose derivative at 0 goes like 
$V_3^\nu$, $\nu\geq 0$, for reasons that will become
clear in Sec.3 below. A simple
modification, which keeps the large-$V_3$ behaviour intact, is given by
\beq\label{2.9}
V_3^{1/3} \to (1+V_3)^{1/3}-1,
\eeq
but the detailed form is not important. If we
now introduce the (non-negative) {\it scale~factor} $a(s)$ by
\beq\label{2.10}
V_3(s) = a^3(s),
\eeq
we can (after suitable rescaling of $s$ and $a(s)$) write the effective action as
\beq\label{2.11}
S^{eff}_{V_4} = \frac{1}{G} \int_0^t ds \;
\left( a(s)\left(\frac{d a(s)}{ds}\right)^2
+ a(s) - \lam a^3 (s)\right),
\eeq
with the understanding that the linear term should be replaced 
using \rf{2.10} and \rf{2.9} for small $a(s)$.
We emphasize again that we have been led to \rf{2.11}
entirely by  ``observation'' and that one can view
the small-$a(s)$ behaviour implied by \rf{2.9}
as a result of quantum fluctuations.

\subsection*{3. Minisuperspace}

\begin{figure}[ht]
\centerline{\scalebox{0.7}{\rotatebox{0}{\includegraphics{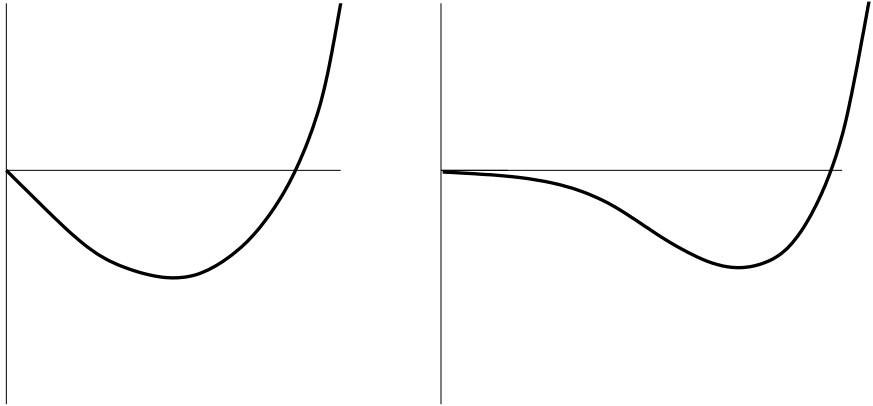}}}}
\caption[phased]{{\small The potential $V(a)$ of \rf{3.3} 
underlying the standard minisuperspace 
dynamics (left)
and the analogous potential in the effective action obtained from the
full quantum gravity model, with small-$a$ modification due to quantum 
fluctuations (right).}}
\label{fig4}
\end{figure}

Let us now consider the simplest minisuperspace model for a
closed universe in quantum cosmology, as 
for instance used by Hartle and Hawking in their semiclassical 
evaluation of the wave function of the universe \cite{hh}.
In Euclidean signature and proper-time coordinates, the
metrics are of the form 
\beq\label{3.1}
ds^2 = dt^2 + a^2(t) d\Omega_3^2,
\eeq
where the scale factor $a(t)$ is the only dynamical variable
and $d\Omega_3^2$ denotes the metric on the three-sphere.
The corresponding Einstein-Hilbert action is
\beq\label{3.2}
S_{eff}= \frac{1}{G} \int dt \left( -a(t)
\left(\frac{d a(t)}{dt}\right)^2
-a(t) + \lam a^3 (t)\right).
\eeq
If no four-volume constraint is imposed, $\lambda$ is the
cosmological constant. If the four-volume is fixed
to $V_4$, such that the discussion parallels the
computer simulations reported above,
$\lam$ should be viewed as a Lagrange multiplier
enforcing a given size of the universe. In the latter
case we obtain the same effective action as in \rf{2.11}
{\it up to an overall sign}, due to the infamous
conformal divergence of the classical Einstein action. 
Let us for the moment ignore this overall minus sign and 
compare the two potentials relevant for the
calculation of semiclassical Euclidean solutions associated 
with the actions \rf{3.2} and \rf{2.11}. The 
``potential''\footnote{To obtain a standard potential -- without changing 
``time" -- one should
first transform to a variable $x=a^{\frac{3}{2}}$ for which
the kinetic term in the actions assumes the standard quadratic
form. It is the resulting
potential $\tilde V(x)=-x^{2/3}+ \lam x^2$ which in the case of 
\rf{2.11} should be modified 
for small $x$ such that $\tilde V'(0)= 0$.} is
\beq\label{3.3}
V(a)= -a + \lam a^3,
\eeq
and is shown in Fig.\ \ref{fig4}, without and with small-$a$ modification, 
for the standard 
minisuperspace model and our effective model, respectively.
The quantum-induced difference for small $a$ is important
since the action  \rf{2.11} allows for a classically stable solution $a(t)=0$ 
which explains the ``stalk'' observed in the computer simulations.
Moreover,  it is appropriate to speak of a Euclidean ``bounce'' because
$a=0$ is a local maximum. If one therefore
{\it naively}  turns the potential upside down
when rotating back to Lorentzian signature, the metastable state 
$a(t)=0$ can tunnel to a state where $a(t) \sim V_4^{1/4}$, 
with a probability
amplitude per unit time which is (the exponential of) the Euclidean
action. We will discuss this further in the next section.

In order to understand how well the semiclassical action 
\rf{2.11} can reproduce the Monte Carlo data, that is, the correlator
$C_{V_4}(\Delta)$ of Fig.\ref{fig2}, we have solved for the
semiclassical bounce using \rf{2.11}, and presented the result
as the continuous black curve in Fig.\ref{fig2}.\footnote{More precisely, 
we solved the classical equation of motion
corresponding to the potential shown in Fig.\ref{fig4} on the right, with an energy
slightly below zero (the closer to zero the longer the stalk), and used
this solution to create an artificial distribution of three-volumes $V_3(s)$
analogous to the one generated by Monte Carlo simulation
from first principles. We then treated this artificial distribution
precisely as if it had come from real Monte Carlo data.}.
The agreement with the real data generated by the
Monte Carlo simulations is clearly perfect.

\subsection*{4. The wave function of the universe}

The picture emerging from the above for the effective dynamics 
of the scale factor resembles that of a universe created by
tunneling from nothing (see, for example, \cite{vilenkin,linde,
rubakov}), although the presence of a preferred notion of
time makes our situation closer to conventional quantum
mechanics. In the set-up analyzed here, there is apparently a
state of vanishing spatial extension which can ``tunnel" to
a universe of finite linear extension of order $a \sim V_4^{1/4}$.
Adopting such a tunneling interpretation, the action 
of the bounce is
\beq\label{4.1}
S_{V_4}^{eff} \sim \frac{V_4^{1/2}}{G},
\eeq
and the associated probability per unit proper 
time for the tunneling given by
\beq\label{4.2}
P(V_4) \sim e^{-S^{eff}_{V_4}}.
\eeq
However, recall that this picture arose from a situation where
for computer-technical reasons
we imposed a constraint on the four-volume. Since
our formulation possesses a well-defined 
Hamiltonian\footnote{In the framework of 
Causal Dynamical Triangulations one has a transfer matrix 
between consecutive proper-time slices whose logarithm 
gives in principle
the full quantum Hamiltonian, see \cite{ajl4d} for details.} 
which is bounded
below, the ground state wave function can be chosen real and {\it positive}.
In view of this, calling \rf{4.2} a tunneling probability is
misleading, since it would imply an oscillating behaviour
for $a\gg V_4^{1/4}$. The correct interpretation of \rf{4.2}
is rather that of the square of the ground state wave function
for $a \sim V_4^{1/4}$.

To illustrate what we have in mind, let us consider a 
quantum-mechanical system with Hamiltonian $H= p^2/2 + V(x)$, 
where the minimum of the potential $V$ is at $x\equ 0$.
The ground state wave function has the path integral representation 
\beq\label{4.10}
\Psi_0 (a) \sim \int_{x(-\infty) = 0}^{x(0)=a} \cD x(t)
\; e^{- S_E[x(t)]},
\eeq
where $S_E[x(t)]$ is the classical Euclidean action
\beq\label{4.11}
S_{E}[x(t)]= \int_{-\infty}^0 dt \; \left[\oh {\dot{x}}^2- (-V(x))\right].
\eeq
If there is a classical solution 
$x_{cl}(t)$ which extremizes the Euclidean action \rf{4.11}
with boundary conditions $x_{cl}(-\infty)=0$ and $x_{cl}(0)=a$, 
a semiclassical
calculation of $\Psi_0(a)$ involves a saddle point expansion around
that solution and the leading exponential of the wave function is
\beq\label{4.12}
\Psi_0(a) \sim e^{-S_{E}[x_{cl}(t)]},~~~~S_E[x_{cl}(t)]=
\int_0^a dx \, \sqrt{2 V(x)}.
\eeq
As an example, for the harmonic oscillator we have 
$x_{cl}(t) = a \, e^{\om t}$ and
\rf{4.12} is exact. For a general potential the semiclassical 
approximation will of course not be exact. Nevertheless, we have presented
strong evidence that in the case of quantum gravity, integrating over 
all degrees of freedom except the three-volume and defining 
$a = V_3^{1/3}$, the semiclassical approximation is excellent. 
If we assume that it is equally good 
in the absence of the four-volume constraint, we compute in a
straightforward way from \rf{4.12} that
\beq\label{4.13}
\Psi_0(a) \sim e^{-\frac{c}{\La G}((1+\La a^2)^{3/2}-1)},
\eeq
with $c$ a constant of order one. Note that $\La$ in \rf{4.13}
is the real cosmological constant and no longer a Lagrange
multiplier. {\it We have thus calculated the wave function of the universe
from first principles} up to prefactors and
corrections to the semiclassical approximation.

It is important to understand that the wave function $\Psi_0(a)$ {\it can} 
be addressed 
via computer simulations using a decomposition analogous to
\rf{2.1a}, namely, 
\beq\label{4.14}
\Psi_0(a) = \int dV_4 \; e^{-\frac{\La}{G} V_4}\; \tilde\Psi_0(a;V_4),
\eeq 
\beq\label{4.15}
\tilde \Psi_0 (a;V_4) = \int_0^a \cD[g] \; e^{\frac{1}{G} \int d^4x \sqrt{\det g}R} \;
\del(\int d^4x \sqrt{\det g} -V_4),
\eeq
where the functional integration in \rf{4.15} is over four-geometries with 
$V_3(-\infty)=0$ and $V_3(t\equ 0)= a^3$. The computer simulations
reported here were done for the special case 
$V_3(t\equ 0)\equ 0$.\footnote{Since the spatial 
volume of the universe is not a monotonic function of $V_4$, 
the integration range in \rf{4.15} in this case should be split into two intervals
$[0,a_{\rm max}]$ and $[a_{\rm max},0]$.}
Using \rf{4.14} and \rf{4.15} one can now check whether the semiclassical 
approximation reported here is valid for all values of $a$ and $V_4$. 
If so, one will be led to \rf{4.13}.

\subsection*{5. Discussion}

Causal Dynamical Triangulations constitute a framework for defining
quantum gravity nonperturbatively as the continuum
limit of a well-defined regularized sum over geometries. 
We reported recently on the outcome of the first 
Monte Carlo simulations in four dimensions \cite{ajl-prl}. 
Very encouragingly, we observed the dynamical generation of a
macroscopic four-dimensional (Euclidean) world, with small quantum
fluctuations superimposed. 
In this letter we showed that the scale factor characterizing the macroscopic
shape of this ground state of geometry is well described by an effective
action similar to that of the simplest minisuperspace
model used in quantum cosmology. However, in our case such
a result has for the
first time -- we believe -- been derived from first principles.

The negative sign of the kinetic term in the standard 
minisuperspace action \rf{3.2} 
reflects the well-known unboundedness of the conformal mode in the 
Euclidean Einstein-Hilbert action. 
Amazingly, after integrating out all variables except the
scale factor (which is simply the global conformal mode), we obtain a 
{\it positive} kinetic term in our effective action \rf{2.11}.
This is consistent with a continuum
formulation of the gravitational path integral in proper-time gauge,
where a strong argument was made for the nonperturbative cancellation 
of the conformal
divergence by a measure factor coming from a Faddeev-Popov
determinant \cite{loll}. Similar ideas were pursued 
in \cite{amm}, although in that case matter fields were needed
to change the sign of the conformal mode term.
The phenomenon of sign change of the
conformal kinetic term is also familiar from the 2d Euclidean 
quantum theory where,
due to the conformal anomaly, integrating out unitary matter
with central charge $0 \leq c \leq 1$ yields an effective
term $-c (\partial \phi)^2$ for the conformal factor $\phi$ in the action,
making it unbounded below. Again it is a Faddeev-Popov determinant,
arising from the requirement to integrate not over metrics, but only
geometries, which adds a $26 (\partial \phi)^2$ and ensures that the
combined kinetic term is positive. 

A number of open issues remain to be addressed,
including the details of the renormalization mechanism.
Here Causal Dynamical Triangulations gives us the possibility to 
study Weinberg's scenario of ``asymptotic safety"
\cite{weinberg} in the context of an explicit quantum-gravitational model.
As indicated in
earlier work on Causal Dynamical Triangulations in space-time
dimension three, the renormalization may be non-standard \cite{ajlabab}, 
which in a way would be welcome.  This is also supported by the present 
computer simulations, in the sense that no fine-tuning of the bare gravitational 
coupling constant seems to be necessary to reach the 
continuum limit. Details of this will be discussed elsewhere \cite{more}. 

A most interesting question is of course how the above semiclassical 
cosmological picture is changed by the inclusion of matter fields.
We have now the chance to investigate a number of possible 
scenarios suggested in quantum cosmology from first principles.

\subsection*{Acknowledgment}
We thank Y. Watabiki for discussions.
J.A.\ and J.J.\ were supported by ``MaPhySto'', 
the Center of Mathematical Physics 
and Stochastics, financed by the 
National Danish Research Foundation.
J.J.\ acknowledges support by the Polish Committee  for Scientific Research 
(KBN) grant 2P03B 09622 (2002-2004).


\begin{thebibliography}{99}

\bibitem{eqg} G.W.\ Gibbons and S.W.\ Hawking: 
{\it Euclidean quantum gravity}, World Scientific, Singapore, 1993.

\bibitem{hh}
J.B.~Hartle and S.W.~Hawking:
{\it Wave function of the universe,}
Phys.\ Rev.\ D\ 28 (1983) 2960-2975.

\bibitem{vilenkin}
A.~Vilenkin:
{\it Creation of universes from nothing},
Phys.\ Lett.\ B\ 117 (1982) 25-28;
{\it Quantum creation of universes},
Phys.\ Rev.\ D 30 (1984) 509-511.

\bibitem{linde}
A.D.~Linde:
{\it Quantum creation of the inflationary universe},
Lett.\ Nuovo Cim.\ 39 (1984) 401-405.

\bibitem{rubakov}
V.A.~Rubakov:
{\it Quantum mechanics in the tunneling universe},
Phys.\ Lett.\ B\ 148 (1984) 280-286.

\bibitem{hl}
J.J.~Halliwell and J.~Louko:
{\it Steepest descent contours in the path integral approach to quantum
cosmology. 1. The de Sitter minisuperspace model},
Phys.\ Rev.\ D\ 39 (1989) 2206-2215.

\bibitem{haha}
J.J.~Halliwell and J.B.~Hartle:
{\it Integration contours for the no boundary wave function 
of the universe},
Phys.\ Rev.\ D\ 41 (1990) 1815-1834.

\bibitem{vilrev}
A.~Vilenkin:
{\it Quantum cosmology and eternal inflation},
in ``The future of theoretical physics and cosmology",
eds. G.W.\ Gibbons, E.P.S.\ Shellard and S.J.\ Rankin, Cambridge 
University Press (2003) 649-666 [gr-qc/0204061].

\bibitem{al} J.\ Ambj\o rn and R.\ Loll:
{\it Non-perturbative Lorentzian quantum gravity, causality and 
topology change},
Nucl.\ Phys.\ B\ 536 (1998) 407-434 [hep-th/9805108].

\bibitem{ajl3d} J.\ Ambj\o rn, J.\ Jurkiewicz and R.\ Loll:
{\it Nonperturbative 3-d Lorentzian quantum gravity},
Phys.\ Rev.\ D\ 64 (2001) 044011 [hep-th/0011276].

\bibitem{ajl-prd} J.\ Ambj\o rn, J.\ Jurkiewicz and R.\ Loll:
{\it A nonperturbative Lorentzian path integral for gravity},
Phys.\ Rev.\ Lett.\ 85 (2000) 924-927 [hep-th/0002050].

\bibitem{ajl4d} J.\ Ambj\o rn, J.\ Jurkiewicz and R.\ Loll:
{\it Dynamically triangulating Lorentzian quantum gravity},
Nucl.\ Phys.\ B\ 610 (2001) 347-382 [hep-th/0105267].

\bibitem{ajl-prl}
J.~Ambj\o rn, J.~Jurkiewicz and R.~Loll:
{\it Emergence of a 4D world from causal quantum gravity},
Phys.\ Rev.\ Lett.\ 93 (2004) 131301 [hep-th/0404156].

\bibitem{teitelboim} C.\ Teitelboim:
{\it Causality versus gauge invariance in quantum gravity and supergravity}, 
Phys.\ Rev.\ Lett.\  50 (1983) 705-708;
{\it The proper time gauge in quantum theory of gravitation},
Phys.\ Rev.\ D28 (1983) 297-309. 

\bibitem{vilcos} A.\ Vilenkin:
{\it Approaches to quantum cosmology},
Phys.\ Rev.\ D\ 50 (1994) 2581-2594 [gr-qc/9403010].

\bibitem{MC} J.\ Ambj\o rn, B.\ Durhuus and T.\ Jonsson: {\em Quantum
geometry}, Cambridge Monographs on Mathematical Physics, 
Cambridge University Press, Cambridge, UK, 1997. 

\bibitem{bark} M.E.J.\ Newman and G.T.\ Barkema: {\it Monte Carlo
methods in statistical physics}, Oxford University Press, Oxford, UK,
1999. 

\bibitem{more} J.~Ambj\o rn, J.~Jurkiewicz and R.~Loll, to appear.

\bibitem{loll}
A.~Dasgupta and R.~Loll:
{\it A proper-time cure for the conformal sickness in quantum gravity},
Nucl.\ Phys.\ B\ 606 (2001) 357-379 [hep-th/0103186].

\bibitem{amm}
I.~Antoniadis, P.O.~Mazur and E.~Mottola:
{\it Conformal symmetry and central charges in four-dimensions},
Nucl.\ Phys.\ B\ 388 (1992) 627-647 [hep-th/9205015].

\bibitem{weinberg}
S.~Weinberg:
{\it Ultraviolet divergences in quantum theories of gravitation},
in ``General relativity: Einstein centenary survey", eds. S.W.\ Hawking 
and W.\ Israel, Cambridge University Press, Cambridge, UK (1979) 790-831.

\bibitem{ajlabab}
J.~Ambj\o rn, J.~Jurkiewicz and R.~Loll:
{\it Renormalization of 3d quantum gravity from matrix models,}
Phys.\ Lett.\ B\ 581 (2004) 255-262 [hep-th/0307263].


\end{thebibliography}
\end{document}